\documentclass[crystals,article,accept,moreauthors,pdftex]{Definitions/mdpi}

\firstpage{1}
\pubvolume{xx}
\issuenum{1}
\articlenumber{5}
\pubyear{2020}
\copyrightyear{2020}
\history{Received: date; Accepted: date; Published: date}
 \updates{yes} 

\usepackage{mathdesign,upgreek}

\Title{Colossal Negative Magnetoresistance Effect in A La$_{1.37}$Sr$_{1.63}$Mn$_2$O$_7$ Single Crystal Grown by Laser-Diode-Heated Floating-Zone Technique} 

\Author{Si Wu $^{1}$, 
Yinghao Zhu $^{1}$, Junchao Xia $^{1}$, Pengfei Zhou $^{1}$, Haiyong Ni $^{2,}$*\orcidA{} and Hai-Feng Li $^{1,}$*\orcidA{}}

\AuthorNames{Si Wu, Yinghao Zhu, Junchao Xia, Pengfei Zhou, Haiyong Ni and Hai-Feng Li}

\address{%
$^{1}$ \quad Joint Key Laboratory of the Ministry of Education, Institute of Applied Physics and Materials Engineering, University of Macau, Avenida da Universidade, Taipa, Macao SAR 999078, China;  yb87826@connect.umac.mo (S.W.); yb87818@connect.um.edu.mo (Y.Z.); yb97833@connect.um.edu.mo (J.X.); yb97828@umac.mo (P.Z.)\\
$^{2}$ \quad Guangdong Province Key Laboratory of Rare Earth Development and Application, Guangdong Research Institute of Rare Metals, Guangdong Academy of Sciences, Guangzhou 510651, China
}

\corres{Correspondence: haifengli@um.edu.mo (H.-F.L.);  Nhygd@163.com (H.N.); Tel.: +853-8822-4035 (H.-F.L.)}

\abstract{We have grown La$_{1.37}$Sr$_{1.63}$Mn$_2$O$_7$ single crystals with a laser-diode-heated floating-zone furnace and studied the crystallinity, structure, and magnetoresistance (MR) effect by in-house X-ray Laue diffraction, X-ray powder diffraction, and resistance measurements. The La$_{1.37}$Sr$_{1.63}$Mn$_2$O$_7$ single crystal crystallizes into a tetragonal structure with space group \emph{I}4{/}\emph{mmm} at room temperature. At 0 T, the maximum resistance centers around $\sim$166.9 K. Below $\sim$35.8 K, it displays an insulating character with an increase in resistance upon cooling. An applied magnetic field of \emph{B}~=~7~T strongly suppresses the resistance indicative of a negative MR effect. The minimum MR value equals $-$91.23\% at 7 T and 128.7 K. The magnetic-field-dependent resistance shows distinct features at 1.67, 140, and 322 K, from which we calculated the corresponding MR values. At 14 T and 140 K, the colossal negative MR value is down to $-$94.04(5)\%. We schematically fit the MR values with different models for an ideal describing of the interesting features of the MR value versus \emph{B} curves.}

\keyword{single crystal; laser-diode-heated floating-zone technique; crystalline; magnetoresistance effect}

\begin{document}

\section{Introduction}

During the past decades, the study of $R_{1-x}A_x$MnO$_3$ materials, where \emph{R} and \emph{A} represent trivalent rare-earth and divalent alkaline-earth ions, respectively, and Mn ions locate at the center of O$_6$ octahedra that build up a three-dimensional network in the way of sharing corners, has been performed intensively. These compounds attract considerable attention because they exhibit metal-insulator transitions, a colossal magnetoresistance (MR) effect, charge/orbital ordering, and other fascinating properties~\cite{Jonker1950, Helmot1993, Chahara1990, Murata2003}. Tailoring the chemical pressure by doping can induce a phase transition from the paramagnetic or antiferromagnetic insulating state to the ferromagnetic metallic one with a thermal change~\cite{Urushibara1995, Schiffer1995, Hwang1995}. Below the ferromagnetic phase transition temperature, ferromagnetic clusters/polarons (i.e., local short-range ferromagnetic spin regimes) form in the whole sample, accompanied by a sharp decrease in resistivity with a discontinuous drop of the equilibrium Mn--O bond length~\cite{Goodenough1997, Teresa1997, Uehara1999, Li2012}. The first-order phase transition from insulating to metallic behavior was confirmed by neutron scattering in the study of La$_{2/3}$Ca$_{1/3}$MnO$_3$ compound~\cite{Teresa1997}.

\begin{figure}[H]
\centering
\includegraphics[width=0.48\textwidth] {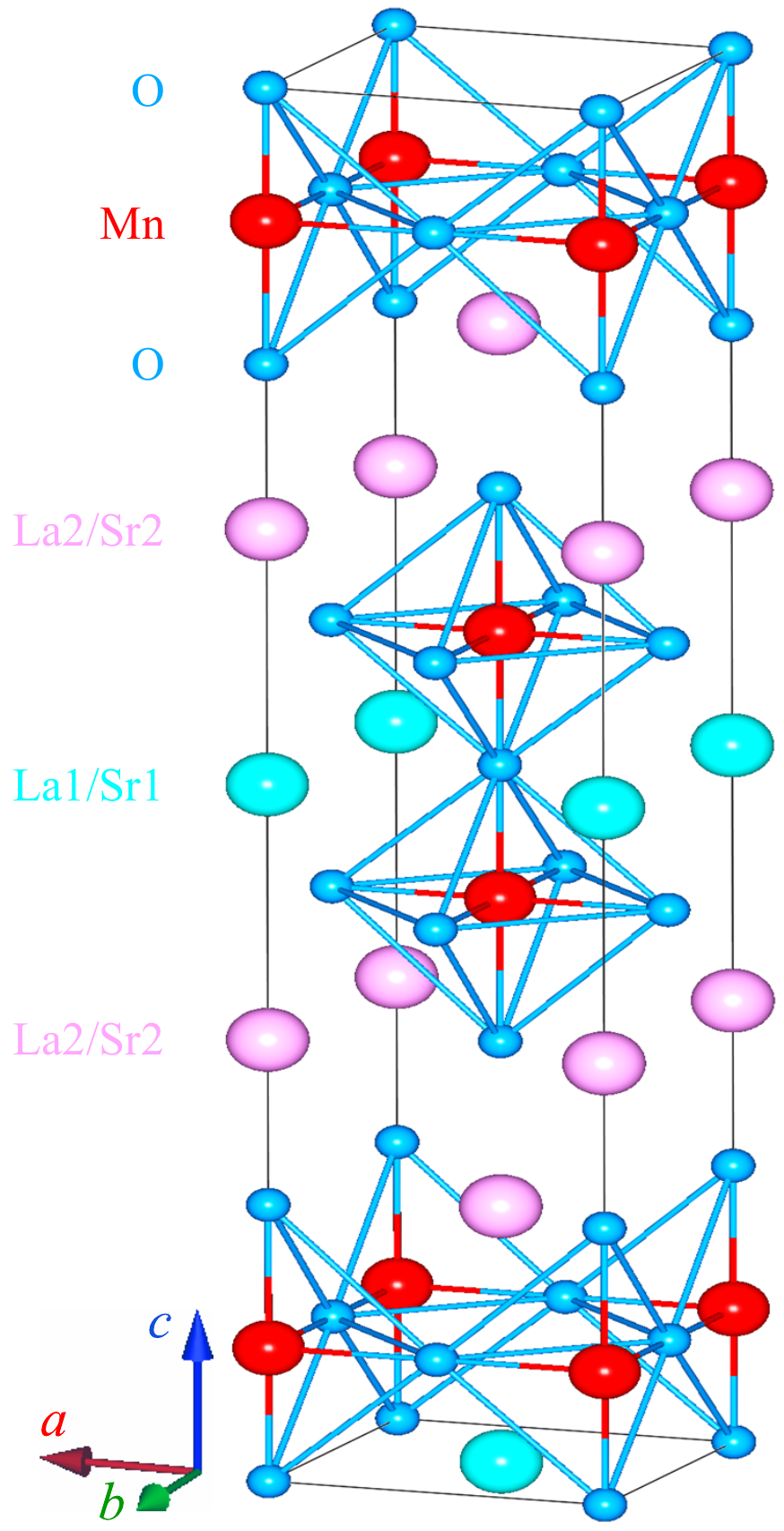}
\caption{(color online)
Tetragonal crystal structure of a La$_{1.37}$Sr$_{1.63}$Mn$_2$O$_7$ single crystal in one unit cell (solid lines) with space group \emph{I}4{/}\emph{mmm} (No. 139). The La, Sr, Mn, and O ions are labeled, and the La and Sr ions are illustrated with the same color code. In this structure, there are two crystallographic sites for La1{/}Sr1 and La2{/}Sr2 ions and three sites for O ions (marked with the same color code).
}
\label{str}
\end{figure}

In 1958, Ruddlesden and Popper~\cite{Ruddlesden1958} synthesized the Sr$_3$Ti$_2$O$_7$ compound with starting materials of SrCO$_3$ and TiO$_2$. It was speculated that every two layers of SrTiO$_3$ would be separated by a layer of SrO~\cite{Ruddlesden1958}. Taking the perovskite-based manganites as one example, the formula of the Ruddlesden--Popper series compounds is generally expressed as ($R_{1-x}A_x)_{n+1}$Mn$_n$O$_{3n+1}$. The Mn and O ions form a MnO$_2$ plane that is arranged in a variety of sequences with MnO$_2$ planes interleaved with ($R, A$)O planes, as shown in Figure~\ref{str}. The annotation of these materials depends on the number of MnO$_2$ planes stacked between ($R, A$)O bilayers~\cite{Seshadri1997, Kubota2000, Li2008}. Compared with the crystal structure of $R_{1-x}A_x$MnO$_3$ compounds within which the MnO$_6$ octahedra form an infinite network of layers in three dimensions, \emph{R}$_{2-2x}A_{1+2x}$Mn$_2$O$_7$ accommodates a quasi-two-dimensional double-layer network. The reduction of dimensionality in these compounds leads to the anisotropic transport property. A metal-insulator phase transition with giant MR effect was reported in the La$_{1.2}$Sr$_{1.8}$Mn$_2$O$_7$ single crystal~\cite{Moritomo1996}. Subsequently, the LaSr$_2$Mn$_2$O$_7$ compound was found to have the complex interplay of ferromagnetic and antiferromagnetic interactions~\cite{Seshadri1997}, leading to interesting electrical transport properties, e.g., a re-entrant insulating phase transition at low temperatures. Due~to the character of low dimensionality, one can easily modify magnetic coupling strengths within the inner- and inter-layers by tailoring chemical pressure~\cite{Battle1996, Moritomo1998, Moritomo1999} and doping concentration~\cite{Asano1997, Hirota1998, Perring1998, Argyriou1999} and even by applying fields~\cite{Perring1998, Moritomo1999}, resulting in complex magnetic structures~\cite{Li2008}.

The fascinating electrical transport property of a Ruddlesden--Popper compound correlates intimately with its detailed magnetic structure~\cite{Battle1996, Moritomo1998, Moritomo1999}. Previously, the crystal structure, transport property, magnetism, magnetoelasticity, and optical properties of La$_{2-2x}$Sr$_{1+2x}$Mn$_2$O$_7$ \mbox{(0.3 $\leq x \leq$ 0.5}) single crystals were intensively studied~\cite{Kimura2000}. The detailed phase diagrams of the magnetic, electrical, and structural properties of La$_{2-2x}$Sr$_{1+2x}$Mn$_2$O$_7$ compounds were compiled~\cite{Qiu2004, Zheng2008}. In order to confirm the relationship between the magnetic structure and the distortion of MnO$_6$ octahedra, the magnetic structure of a La$_{1.37}$Sr$_{1.63}$Mn$_2$O$_7$ single crystal was explored under hydrostatic pressures of up to 0.8 GPa by a neutron diffraction study. The result shows that the ground state does not change under the hydrostatic pressure of up to 0.8 GPa~\cite{Sonomura2013}. This indicates that the magnetic structure depends not only on the energy stability of the crystal field, but also on other interactions, such as magnetic dipole--dipole interactions.

Achieving a full understanding of the coupling between charge, spin, lattice, and orbital degrees of freedom and shedding light on the interesting properties of La$_{2-2x}$Sr$_{1+2x}$Mn$_2$O$_7$ compounds necessitate the growth of high-quality and large single-crystalline samples. In this paper, we have grown single crystals of the La$_{1.37}$Sr$_{1.63}$Mn$_2$O$_7$ compound with a laser-diode-heated floating-zone furnace. We characterized the single crystals with in-house X-ray Laue diffraction, X-ray powder diffraction, and resistance measurements.

\section{Results and Discussion}

\subsection{Single Crystal Growth}

There are several techniques for single crystal growth, for example, floating-zone, Czochralski, Bridgman, top seeded solution, and gas-phase growth~\cite{Li2008}. Among them, the floating-zone method doesn't use a crucible, and the seed and feed rods are freely standing. Thus, the grown crystals should have the highest purity~\cite{Li2008}. Therefore, the floating-zone technique has been widely used to grow single crystals of oxides like La$_{2-2x}$Sr$_{1+2x}$Mn$_2$O$_7$ manganites~\cite{Guptasarma2005}. Compared with the traditional floating-zone furnace with IR-heating halogen lamps, the laser-diode-heated floating-zone furnace holds a higher maximum temperature and a steeper temperature gradient at the liquid--solid interface~\cite{Ito2013}. These advantages are more favorable during the process of single crystal growth~\cite{Li2008, Wu2020}. In this study, the laser-diode-heated floating-zone furnace is equipped with five laser diodes with a wavelength of $\sim$975 nm and a power of $\sim$200 W each. The shape of the laser beam is 4 mm (width) $\times$ 8 mm (height), and the working distances are $\sim$135 mm~\cite{Wu2020}.

\begin{figure}[H]
\centering
\includegraphics[width=0.72\textwidth] {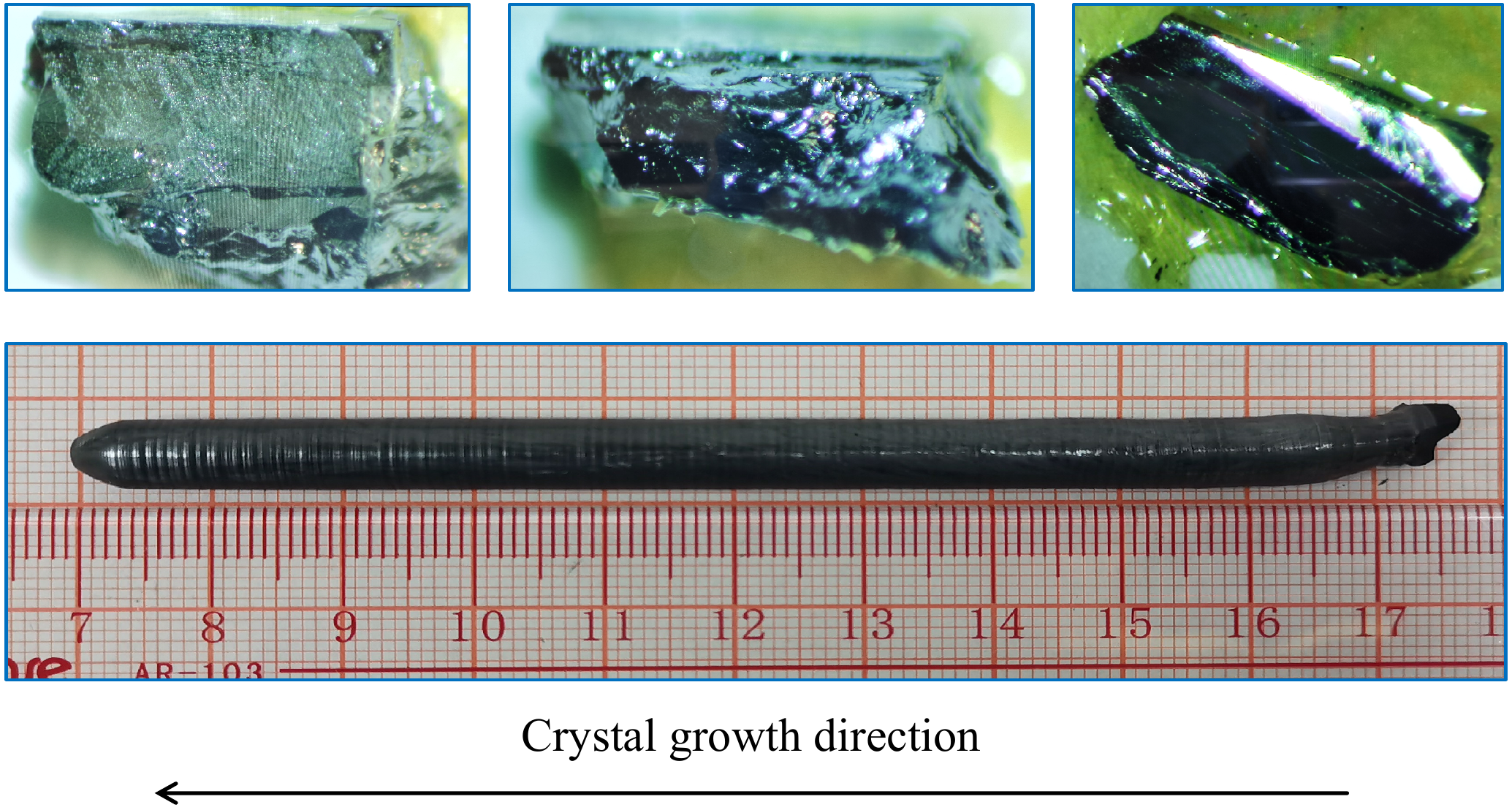}
\caption{(color online) (Top panel) Some cleaved pieces of the grown crystal display the natural crystallographic $ab$ plane and shining surfaces. (Bottom panel) One representative image of a La$_{1.37}$Sr$_{1.63}$Mn$_2$O$_7$ crystal as grown by our laser-diode-heated floating-zone furnace with a mixture of working gases  ($\sim$90\% Ar + $\sim$10\% O$_2$) under a pressure of $\sim$0.5 MPa. The bottom arrow points out the crystal growth direction.}
\label{crystal}
\end{figure}

\begin{figure}[H]
\centering
\includegraphics[width=0.72\textwidth] {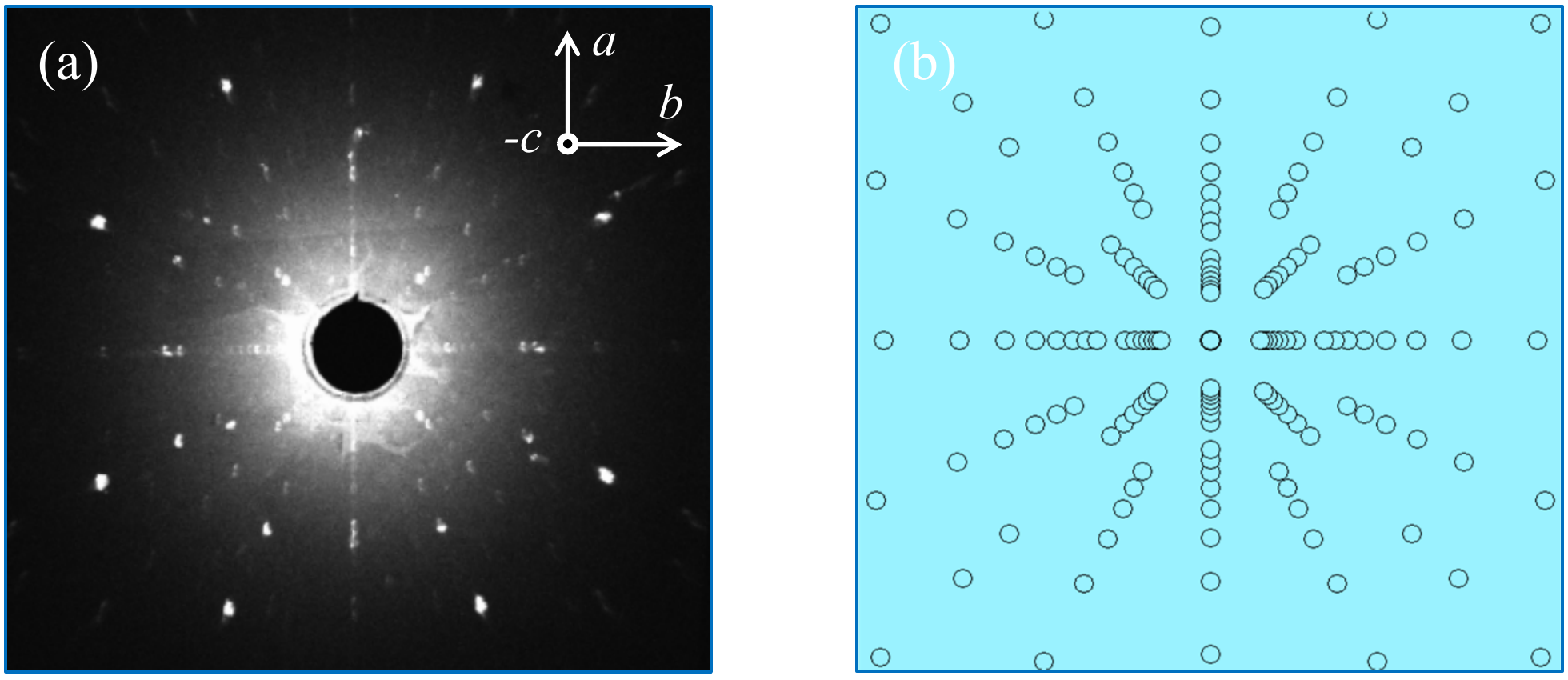}
\caption{(color online)
(\textbf{a}) One representative in-house X-ray Laue pattern of a La$_{1.37}$Sr$_{1.63}$Mn$_2$O$_7$ single crystal with real-space lattice vectors  marked. The incoming X-ray beam is perpendicular to the plane of the Laue spots and thus is along the unit-cell \emph{c} axis. (\textbf{b}) Corresponding simulation of the X-ray Laue pattern with the software ``OrientExpress''~\cite{Ouladdiaf2006}.}
\label{laue}
\end{figure}

During the processes of calcination and sintering, we found that the wall and lid of the alumina crucible became black. After each crystal growth, some black powder was attached to the wall of the quartz. Based on our experience, we infer that the Mn- and Sr-based oxides are constituents of the volatiles. Therefore, we added 5\% more MnO$_2$ and 3\% more SrCO$_3$ oxides for the initial mixture of raw materials. During the initial attempts to grow the La$_{1.37}$Sr$_{1.63}$Mn$_2$O$_7$ single crystal, we adjusted the composition of the working gases. When the content of oxygen was larger than that of argon, we could obtain a stable floating zone and, thus, a steady growth. However, the resultant crystal was comprised of many small grains and was not a single crystal, and our X-ray powder diffraction (XRPD) study demonstrated that it was not a single phase. After many tentative growths, we optimized the growth parameters as: (i) the growth atmosphere had the working gases of $\sim$90\% Ar plus $\sim$10\% O$_2$; (ii) the pressure of the working gases was approximately 0.5 MPa; (iii) the seed and feed rods rotated oppositely at 28 and 30~rpm, respectively; (iv) the stable growth rate was fixed at 4 mm/h; (v) the down speed of the feed rod was kept at 5~mm/h so as to supply sufficient material to the molten zone. Under these conditions, we can grow single crystals of the La$_{1.37}$Sr$_{1.63}$Mn$_2$O$_7$ compound with a diameter of 4--6 mm. The image of one representative La$_{1.37}$Sr$_{1.63}$Mn$_2$O$_7$ crystal as grown is shown in Figure~\ref{crystal} (bottom panel). We cleaved it and found some single crystals with shining natural surfaces, as shown in the top panel of Figure~\ref{crystal}, where the left and right single crystals hold a natural crystallographic $ab$ plane. From the right single crystal, we clearly observed a layer-by-layer growth mode. Figure~\ref{laue}a shows the corresponding in-house X-ray Laue pattern of the flat surface. We simulated the laue pattern with the software ``OrientExpress''~\cite{Ouladdiaf2006}, as shown in Figure~\ref{laue}b, and confirmed that it was a single crystal. The measured surface corresponds to the crystallographic \emph{ab} plane.

\begin{figure}[H]
\centering
\includegraphics[width=0.58\textwidth] {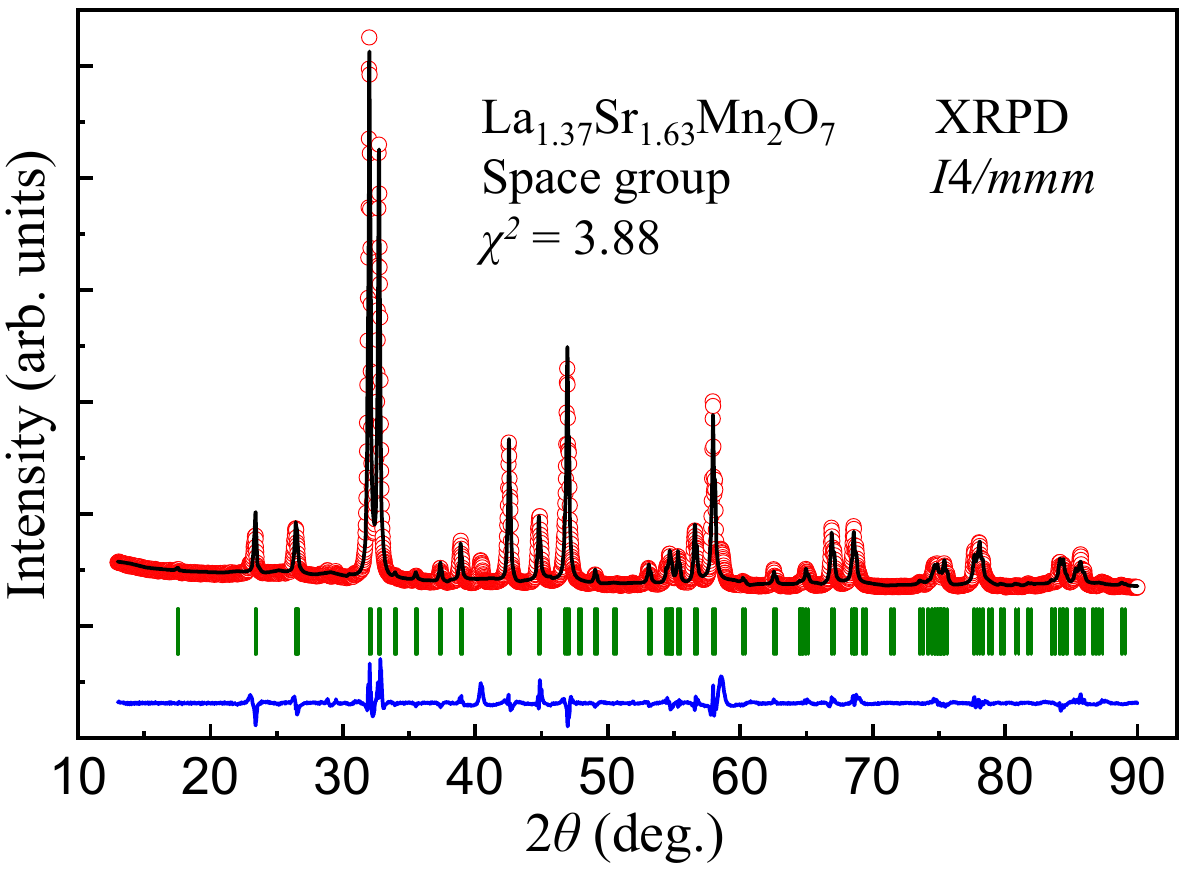}
\caption{(color online)
Observed (circles) and calculated (solid line) room-temperature XRPD patterns collected with a pulverized La$_{1.37}$Sr$_{1.63}$Mn$_2$O$_7$ single crystal at room temperature and refined using the FULLPROF SUITE~\cite{Fullprof}. Vertical bars mark the positions of Bragg peaks. The bottom curve represents the difference between observed and refined XRPD patterns. It is pointed out that the two small extra peaks located at 2$\theta =$ 28.6{--}29.6$^\circ$ were attributed to the radiation contamination from the copper $K_{\beta}$ wavelength. We could not identify the additional peak appearing at 2$\theta =$ 40{--}40.8$^\circ$.}
\label{XRD}
\end{figure}

\subsection{Structure Study}

To check the phase purity and extract structural information of the La$_{1.37}$Sr$_{1.63}$Mn$_2$O$_7$ single crystal, we pulverized a small piece and carried out a room-temperature XRPD experiment. Figure~\ref{XRD} shows the observed and calculated XRPD patterns. We used the software  FULLPROF SUITE~\cite{Fullprof} to refine the collected XRPD pattern. A linear interpolation between automatically selected data points was used to determine the background contribution. We selected the Pseudo-Voigt function to simulate the shape of Bragg peaks. Finally, we refined together the scale factor, zero shift, peak-shape parameters, asymmetry, lattice parameters, atomic positions, and isotropic thermal parameters. Within the present experimental accuracy, the collected Bragg reflections were well indexed with the space group \emph{I}4{/}\emph{mmm}. The low values of goodness of fit indicate a good structural refinement. The refined values of the parameters are listed in Table~\ref{parameters}. The extracted crystal structure in one unit-cell is displayed in Figure~\ref{str}. Within this structure, the Mn and O ions form two layers of corner-shared MnO$_6$ octahedra, and the bilayers of MnO$_6$ octahedra are separated by an insulating La1/Sr1-O layer. There is no rotation of the MnO$_6$ octahedra, keeping the bond angle of Mn--O--Mn at 180$^\circ$. Our XRPD study confirms the main phase of the La$_{1.37}$Sr$_{1.63}$Mn$_2$O$_7$ compound.

\subsection{Scanning Electronic Microscopy}

\begin{figure}[H]
\centering \includegraphics[width=0.58\textwidth]{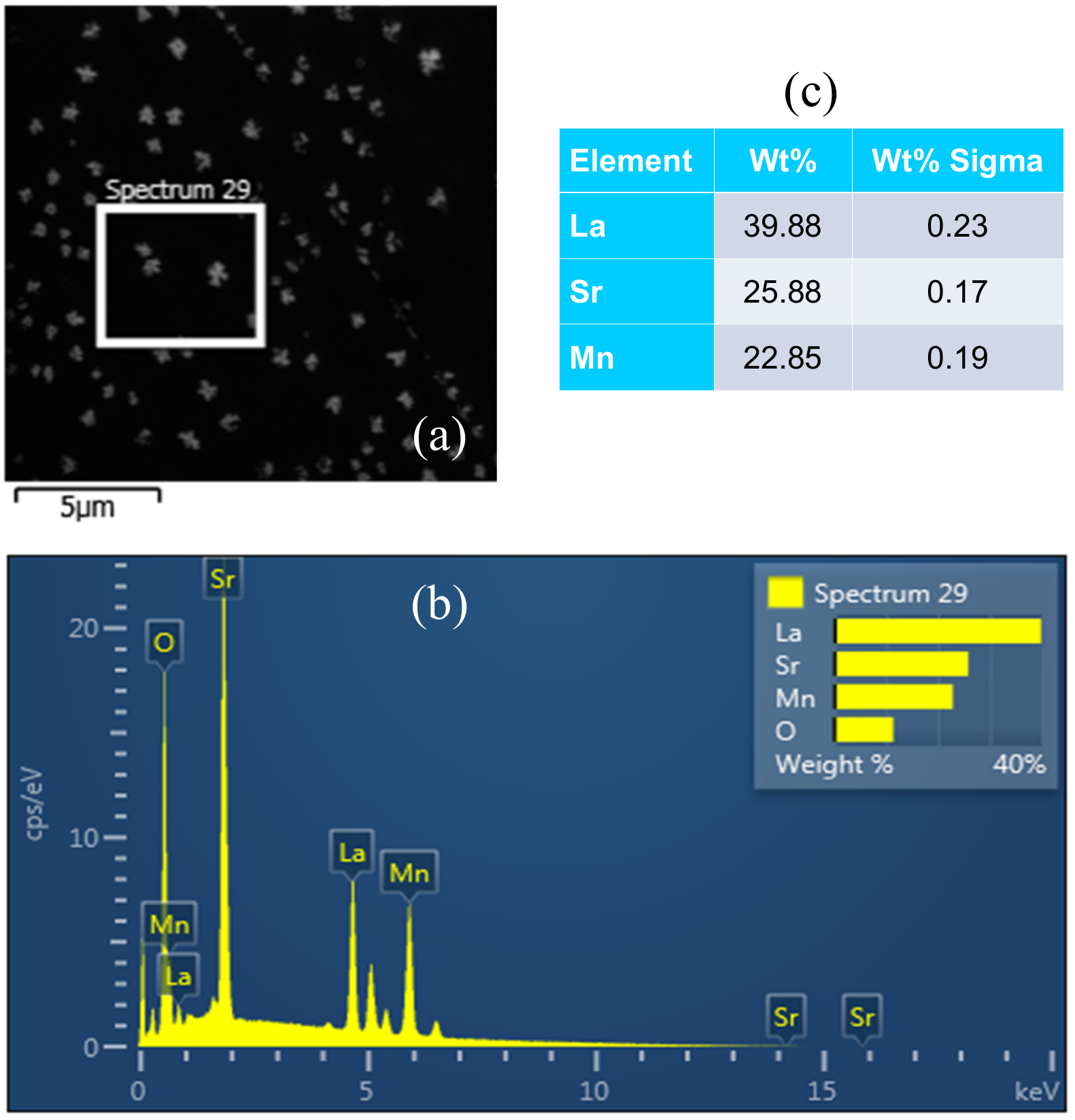}
\caption{(\textbf{a}) Scanning electronic microscopy image of a cleaved La$_{1.37}$Sr$_{1.63}$Mn$_2$O$_7$ single crystal. The scale bar represents 5 $\upmu$m. (\textbf{b}) The energy-dispersive X-ray chemical composition analysis of the selected area that is marked in (a) (the square). (\textbf{c}) Extracted weight percentage (wt. \%) of the chemical compositions of La, Sr, and Mn elements as well as the corresponding error bars (wt. \% sigma). We ignored the composition analysis of oxygen.}
\label{SEMchenfen}
\end{figure}

We randomly selected a piece of the La$_{1.37}$Sr$_{1.63}$Mn$_2$O$_7$ single crystal and studied its chemical composition by the opinion of scanning electronic microscopy, i.e., energy-dispersive X-ray chemical composition analysis. We chose a flat surface with an area $\sim$5 $\times$ 5 $\upmu$m (the square regime as shown in Figure~\ref{SEMchenfen}a) for the detailed statistical study. The corresponding energy-dispersive X-ray spectrum is shown in Figure~\ref{SEMchenfen}b, and the chemical compositions of La, Sr, and Mn elements are listed in Figure~\ref{SEMchenfen}c. We thus extracted a chemical formula, La$_{1.37(1)}$Sr$_{1.41(1)}$Mn$_{1.98(2)}$O$_7$, supposing that the oxygen content is stoichiometric. Clearly, there exist vacancies on the Sr site.

\begin{table}[H]
\caption{Refined room-temperature structural parameters, including lattice constants, unit-cell volume (\emph{V}), atomic positions, isotropic thermal parameters (ITPs), and goodness of refinement, from an in-house XRPD study with a pulverized La$_{1.37}$Sr$_{1.63}$Mn$_2$O$_7$ single crystal. The Wyckoff sites of all atoms were listed. We kept the atomic occupation factors during FULLPROF refinements. We constrained the ITPs of La and Sr ions to being the same, as well as for the O1, O2, and O3 ions. The numbers in parentheses are the estimated standard deviations of the last significant digit.}
\centering
\label{parameters}
\begin{tabular} {llllllllllllllllll}
\toprule
\multicolumn{18}{c} {\textbf{XRPD study of a pulverized La$_{1.37}$Sr$_{1.63}$Mn$_2$O$_7$ single crystal}}                                                                \\
\multicolumn{18}{c} {\textbf{(Tetragonal, Space Group \emph{I}4{/}\emph{mmm} (No. 139), $Z~=~2$)}}                                                                        \\
\midrule
\emph{a}( = \emph{b}) (\AA)                &&&&& \emph{c} (\AA)    &&& \emph{V} ({\AA}$^3$)   &&& \multicolumn{7}{l} {$\alpha ( = \beta~=~\gamma)$ ($^{\circ}$)}          \\
3.8671(1)                                 &&&&& 20.2108(8)        &&& 302.25(2)              &&& \multicolumn{7}{l} {90}                                                  \\
\midrule
Atom              &&&&& Site              &&& \emph{x}      &&& \emph{y}          &&& \emph{z}           &&& ITP                                                           \\
La1/Sr1           &&&&& 2\emph{b}         &&& 0.00          &&& 0.00              &&& 0.50               &&& 2.41(4)                                                       \\
La2/Sr2           &&&&& 4\emph{e}         &&& 0.00          &&& 0.00              &&& 0.3174(1)          &&& 2.41(4)                                                        \\
Mn                &&&&& 4\emph{e}         &&& 0.00          &&& 0.00              &&& 0.0965(2)          &&& 1.88(8)                                                         \\
O1                &&&&& 2\emph{a}         &&& 0.00          &&& 0.00              &&& 0.00               &&& 2.2(1)                                                         \\
O2                &&&&& 4\emph{e}         &&& 0.00          &&& 0.00              &&& 0.1971(7)          &&& 2.2(1)                                                         \\
O3                &&&&& 8\emph{g}         &&& 0.00          &&& 0.50              &&& 0.0958(5)          &&& 2.2(1)                                                     \\
\midrule
\multicolumn{18}{l} {$R_\textrm{p}$~=~5.73, $R_\textrm{wp}$~=~9.20, $R_\textrm{exp}$~=~4.67, and $\chi^2$~=~3.88}                                                       \\
\bottomrule
\end{tabular}
\end{table}

\begin{figure}[H]
\centering
\includegraphics[width=0.60\textwidth] {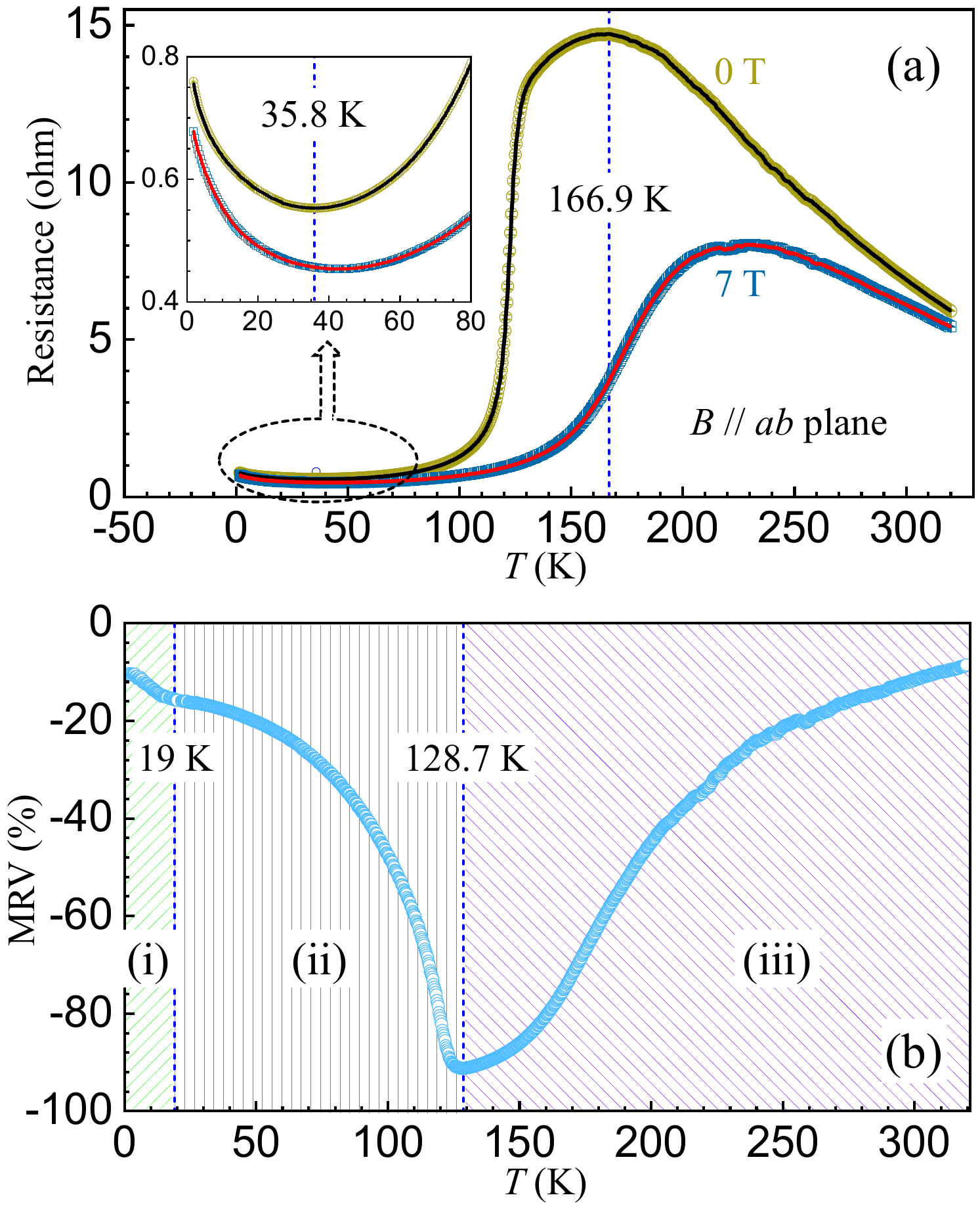}
\caption{(color online) (\textbf{a}) Measured resistance (circles) within the crystallographic \emph{ab} plane of a La$_{1.37}$Sr$_{1.63}$Mn$_2$O$_7$ single crystal with applied magnetic fields \emph{B} (0 and 7 T) perpendicular to the electric current direction in a temperature range from 1.67 to 322 K. Inset is an enlarged image of the data in the temperature range of 1.67--80 K. Error bars are standard deviations, and some are embedded into the symbols. The solid lines are fits to the data (see details in the text).
(\textbf{b}) calculated MRV as a function of temperature at 7 T ( see detailed analysis in the text).}
\label{RT}
\end{figure}

\subsection{Magnetoresistance versus Temperature}

Figure~\ref{RT}a shows the measured resistance of the La$_{1.37}$Sr$_{1.63}$Mn$_2$O$_7$ single crystal within the crystallographic \emph{ab} plane (perpendicular to the applied magnetic-field direction) as a function of temperature at 0 and 7 T. Upon cooling from 322 K, when \emph{B}~=~0 T, the resistance almost increases linearly until \emph{T}~=~166.9 K, where the maximum value appears, centered with a very broad peak. This~temperature point was pushed to \emph{T}~=~229.5 K when \emph{B}~=~7 T. Subsequently, the resistance experiences a sharp decrease from $\sim$128 K. It is worth noting that below 35.8 K, as shown in the inset of Figure~\ref{RT}a, the resistance increases sharply. We define the electrical nature as insulating when $\frac{dR}{dT} < 0$ and conducting when $\frac{dR}{dT} > 0$. Therefore, at \emph{B}~=~0 T and within the studied temperature range, the La$_{1.37}$Sr$_{1.63}$Mn$_2$O$_7$ compound undergoes two electrical phase transitions: insulator (1.67--35.8 K) $\rightarrow$ metal (35.8--166.9~K) $\rightarrow$ insulator (166.9--322 K). When the applied magnetic field \emph{B} equals 7 T, the measured resistance was strongly suppressed, especially in the temperature range within which the maximum appears. When~\emph{B}~=~7~T, the temperature regimes of the electrical phase transitions change into 1.67--43.2 K (insulator), 43.2--229.5~K (metal), and 229.5--322 K (insulator). The suppression and the shift of the measured resistance with a change in the applied magnetic field indicate a strong MR effect.

To quantitatively display the MR effect, we first treated the measured original data (shown as circles in Figure~\ref{RT}a) by manipulating them into evenly-spaced, temperature-dependent points (shown as solid lines in Figure~\ref{RT}a). The manipulated data points agree quite well with the measured data. We can thus calculate the MR value (MRV) by
\begin{eqnarray}
\textrm{MRV}(B, T)~=~\frac{\rho(B, T) - \rho(0, T)}{\rho(0, T)} \times 100\%,
\label{MR}
\end{eqnarray}
where $\rho(B, T)$ and $\rho(0, T)$ are the measured resistances with and without applied magnetic field \emph{B} at a given temperature \emph{T}. With Equation~(\ref{MR}), the calculated MRVs versus temperature are shown in Figure~\ref{RT}b. The colossal negative MRV~=~$-$91.23\% when \emph{T}~=~128.7 K. It~is pointed out that the theoretical minimum MRV~=~$-$100\% when the resistance is suppressed completely by an applied magnetic field. Based on the observed features of the MRV-\emph{T} curve, we can divide it into three regimes upon warming: (i) 1.67--19 K, the negative MRV decreases from $\sim$$-$10.28\% (at 1.67~K) with a kink appearing at $\sim$19 K; (ii) 19--128.7 K, the MRV decreases smoothly, followed by a sharp decrease from $\sim$88 K until the appearance of the minimum MRV at 128.7 K; (iii) 128.7--322~K, the calculated MRV increases step by step. At 322 K, the MRV is $\sim$$-$8.55\%.

\subsection{Magnetoresistance versus Applied Magnetic Field}

As shown in Figure~\ref{RH}a--c, we measured the applied magnetic-field-dependent resistance at 1.67, 140, and 322 K, respectively. At 1.67 K, around 0 T, the resistance $\rho-B$ curve shows a sharp arrow-like shape, i.e., above 0 T, the resistance first sharply and then smoothly decreases, displaying a notching curve; at 140 K, it exhibits a bell-like curve; in contrast, at 322 K, the curve shows a blunt arrow-like shape, i.e., an upper convex curve. These distinct trends imply different microscopic origins of the electrical behaviors.

Based on the following equation,
\begin{eqnarray}
\textrm{MRV}(B)~=~\frac{\rho(B) - \rho(B~=~0)}{\rho(B~=~0)} \times 100\%,
\label{MRB}
\end{eqnarray}
where $\rho(B)$ and $\rho(B~=~0)$ are the measured resistances at applied magnetic fields of \emph{B} and 0 T, respectively, we calculated the corresponding MRVs as a function of the applied magnetic field, as shown in Figure~\ref{RH}a'--c'. We first tried to fit the data to
\begin{eqnarray}
\textrm{MRV}(B)~=~\textrm{k} B^{\alpha},
\label{MR13}
\end{eqnarray}
where k and $\alpha$ are constants. We can fit the data well at 1.67 and 322 K, shown as the solid lines in Figure~{\ref{RH}a',c', respectively. This results in (i) at 1.67 K, k $=$ $-$3.5184(9), and $\alpha =$ 0.6962(2); (ii) at 322 K, k $=$ $-$0.8808(4), and $\alpha =$ 1.2429(2). For the 140 K data at 0--4 T (Figure~\ref{RH}b'), we can fit the calculated MRVs only with
\begin{eqnarray}
\textrm{MRV}(B)~=~\textrm{L}B + \textrm{M}B^2 + \textrm{N}B^3 + \textrm{O}B^4 + \textrm{P}B^5,
\label{MR2}
\end{eqnarray}
with L~=~$-$5.7(3), M~=~7.5(6), N~=~$-$17.8(4), O~=~6.5(1), and P~=~$-$0.68(1). Above 4 T, the MRV approaches a stable state little by little. At 14 T, the negative MRV~=~$-$94.04(5)\%. Normally, the applied magnetic field will increase the resistance of a conductor. Here, we observed a colossal negative MR effect in a La$_{1.37}$Sr$_{1.63}$Mn$_2$O$_7$ single crystal and distinct temperature- and magnetic-field-dependent behaviors.

\begin{figure}[H]
\centering
\includegraphics[width=0.88\textwidth] {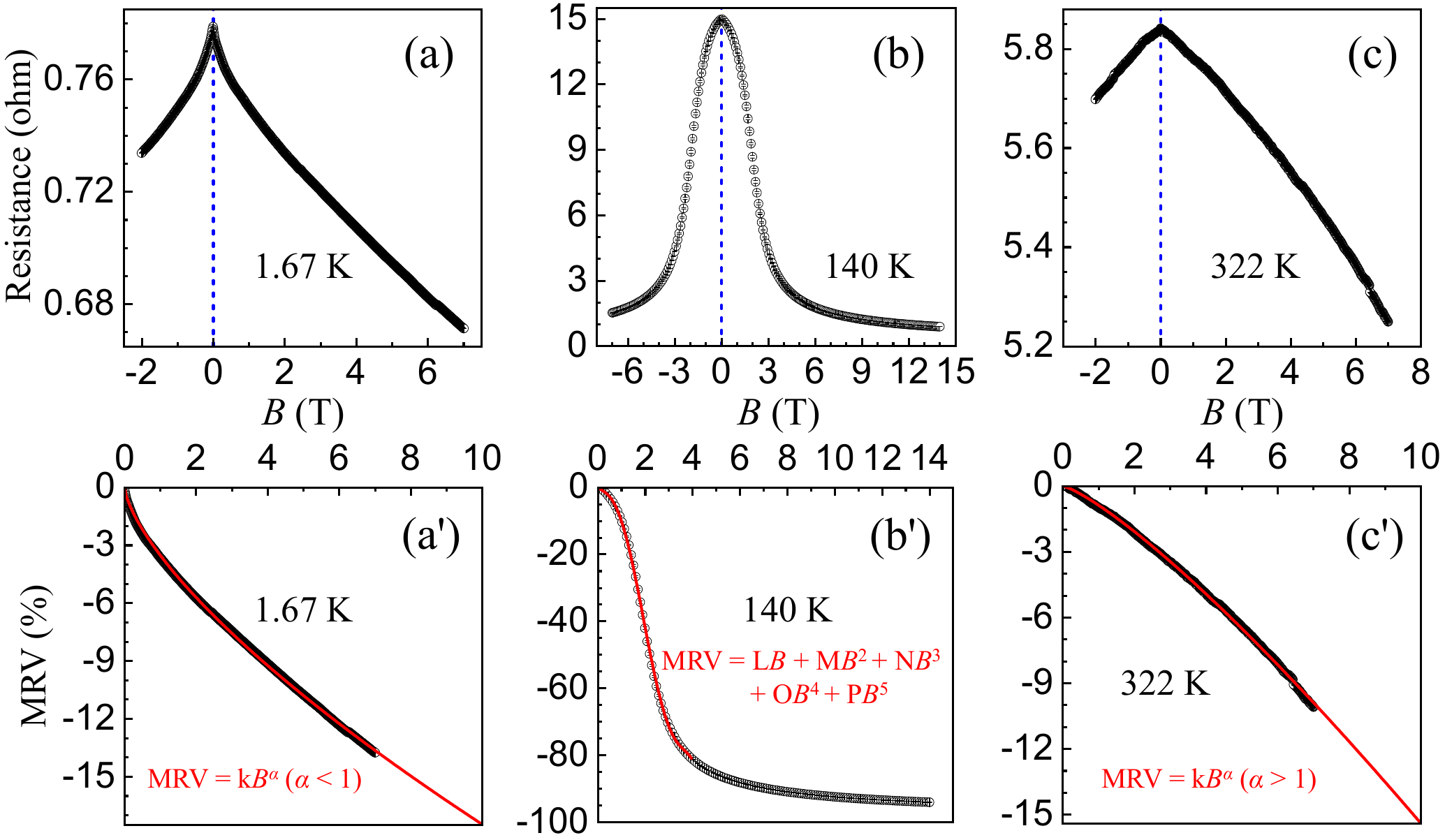}
\caption{(color online) Applied-magnetic-field-dependence of resistance within the crystallographic \emph{ab} plane ($\perp$ \emph{B}) of a La$_{1.37}$Sr$_{1.63}$Mn$_2$O$_7$ single crystal at 1.67 K (\textbf{a}), 140 K (\textbf{b}), and 322 K (\textbf{c}). Error bars are standard deviations and are embedded into the symbols.
(\textbf{a'}--\textbf{c'}): the calculated MRV (circles) versus the applied magnetic field \emph{B}, corresponding to figures (\textbf{a}--\textbf{c}), respectively. The solid lines represent theoretical fits; see the detailed analyses in the text. Error bars were estimated based on the law of propagation of errors, and some were embedded into the symbols.}
\label{RH}
\end{figure}

\section{Materials and Methods}

\subsection{Polycrystal Synthesis}

Polycrystalline samples of the La$_{1.37}$Sr$_{1.63}$Mn$_2$O$_7$ compound were prepared by the traditional solid-state reaction method~\cite{Li2008, Li2007-2}. We chose raw materials of SrCO$_3$ (Alfa Aesar (China) Chemical Co., Ltd, 99.99\%), La$_2$O$_3$ (Alfa Aesar (China) Chemical Co., Ltd, 99.99\%), and MnO$_2$ (Alfa Aesar (China) Chemical Co., Ltd, 99.99\%). These raw materials were kept in a drying oven at 493~K for 48 h, and then quickly weighed at $\sim$473 K to reduce a possible absorption of moisture in the air. The raw materials were well mixed and milled by a Vibratory Micro Mill (FRITSCH PULVERISETTE 0, Fritsch GmbH, Germany) for 1 h. We used alumina crucibles to calcine the mixture of raw materials in air at three temperatures, i.e., 1373 K , 1473 K, and 1623 K, for 48 h each with intermediate grindings. The heating-up and cooling-down ramps were kept at 473~K/h.

\subsection{Preparation of Single Crystal Growth}

After the final grinding, we filled the polycrystalline powder into two plastic cylindrical balloons for seed and feed rod preparation. Both rods were hardened with a hydrostatic pressure of $\sim$70 MPa for 20 min. The shaped rods were sintered at 1573 K for 36 h in the air. Finally, homogeneously densified seed and feed rods with a homogeneous composition distribution were obtained. We have grown bulk single crystals of a La$_{1.37}$Sr$_{1.63}$Mn$_2$O$_7$ compound with a laser-diode-heated floating-zone furnace (Model: LD-FZ-5-200W-VPO-4 PC-UM, Crystal Systems Corporation, Japan)~\cite{Zhu2020, Wu2020}.

\subsection{In-House Characterizations}

Some small pieces of the grown crystals were ground into powder and then characterized by XRPD from 2$\theta$~=~10 to 90$^\circ$ with a step size of 0.02$^\circ$ on an in-house diffractometer employing copper $K_{\alpha1}$ and $K_{\alpha2}$ with a ratio of 2{:}1 as the radiation with a voltage of 45 kV, a current of 200 mA, and ambient conditions. We obtained  single crystals with a shining natural crystalline face (i.e., regular crystallographic \emph{ab} plane). We carried out an in-house X-ray Laue diffraction study to determine the crystallographic orientations. Temperature- and applied-magnetic-field-dependent resistances within the crystallographic \emph{ab} plane were measured on a Quantum Design Physical Property Measurement System (PPMS DynaCool instrument).

\section{Conclusions}

To summarize, we have grown  La$_{1.37}$Sr$_{1.63}$Mn$_2$O$_7$ single crystals with a laser-diode-heated floating-zone furnace. The cleaved single crystal has a shining natural crystallographic face. The~single crystal was grown with $\sim$0.5 MPa working gases of ($\sim$90\% Ar + $\sim$10\% O$_2$), opposite rotations of the seed (28 rpm) and feed (30 rpm) rods, and a growth speed of 4 mm/h. An in-house X-ray Laue study shows regularly arranged diffraction spots, a characteristic of single-crystalline samples. We determined that the shining surface is the crystallographic \emph{ab} plane. The room-temperature XRPD pattern can be well fitted by a tetragonal structure with space group \emph{I}4{/}\emph{mmm}. We extracted  structural information such as lattice constants, unit-cell volume, and atomic positions. The temperature-dependent resistance displays three distinct regimes with different electrical conducting behaviors, i.e., insulator (1.67--35.8~K), metal (35.8--166.9 K), and insulator (166.9--322 K). The applied magnetic field strongly suppresses the resistance, leading to a negative MR effect with a minimum value of $-$91.23\% at \emph{T}~=~128.7 K. By fitting the magnetic-field-dependent resistance, we reveal distinct behaviors, i.e., the 1.67 and 322 K data can be well fitted by Equation~(\ref{MR13}) with $\alpha_{\textrm{1.67K}} =$ 0.6962(2) ($<$ 1) and $\alpha_{\textrm{322K}} =$ 1.2429(2) ($>$ 1); in contrast, the 140 K data at 0--4 T can only be fitted with a quintic Equation~(\ref{MR2}), indicating different microscopic origins.

Presently, we are still working on how to improve the quality and mass of the La$_{1.37}$Sr$_{1.63}$Mn$_2$O$_7$ single crystals, which is required for further studies.
\vspace{6pt}

\authorcontributions{S.W. prepared the materials and grew the single crystals. S.W., Y.Z., J.X., and P.Z. performed experiments and characterized the samples.
S.W. and H.-F.L. analyzed the data and made the figures. All authors discussed and analyzed the results. S.W. and H.-F.L. wrote the main manuscript text. All authors commented on the manuscript and reviewed the paper. H.N. and H.-F.L. conceived and directed the project. All authors have read and agreed to the published version of the manuscript.}

\funding{This project was funded by the University of Macau (File no. SRG2016{-}00091{-}FST), the Science and Technology Development Fund, Macao SAR (Files No. 063/2016/A2, No. 064/2016/A2, No. 028/2017/A1, and No. 0051/2019/AFJ), the Guangdong--Hong Kong--Macao Joint Laboratory for Neutron Scattering Science and Technology, and the Guangdong Science and Technology Project (2019gdasyl-0502005).}

\conflictsofinterest{The authors declare no conflict of interest.}

\reftitle{References}


\begin{thebibliography}{-------}
\providecommand{\natexlab}[1]{#1}

\end{thebibliography}


\begin{thebibliography}{999}

\bibitem{Jonker1950} Jonker, G.H.; Santen, J.H.V. Ferromagnetic compounds of manganese with perovskite structure. \emph{Physica} \textbf{1950}, \emph{16}, 337.
\bibitem{Helmot1993} Helmot, R.V.; Wecker, J.; Holzapfel, B.; Schultz, L.; Samwer, K. Giant Negative Magnetoresistance in Perovskitelike La$_{2/3}$Ba$_{1/3}$MnO$_x$ Ferromagnetic Films. \emph{Phys. Rev. Lett.} \textbf{1993}, \emph{71}, 2331.
\bibitem{Chahara1990} Chahara, K.; Ohno, T.; Kasai, M.; Kozono, Y. Magnetoresistance in magic manganese oxide with intrinsic antiferromagic spin structure. \emph{Appl. Phys. Lett.} \textbf{1993}, \emph{63}, 1990.
\bibitem{Murata2003} Murata, T.; Terai, T.; Fukuda, T.; Kakeshita, T.; Kishio, K. Influnce of Grain Boundray on Magnetoresistance in Hole Doped Manganites La$_{0.7}$Ca$_{0.3}$MnO$_3 $ and (La$_{0.75}$Y$_{0.25}$)$_{0.7}$Sr$_{0.3}$MnO$_3$. \emph{Trans. Mater. Res. Soc. Jpn.} \textbf{2003}, \emph{44}, 2589.
\bibitem{Urushibara1995} Urushibara, A.; Moritomo, Y.; Arima, T.; Asamitsu, A.; Kido, G.; Tokura, Y. Insulator-Metal Transition and Giant Magnetoresistance in La$_{1-\emph{x}}$Sr$_\emph{x}$MnO$_3$. \emph{Phys. Rev. B} \textbf{1995}, \emph{51}, 14103.
\bibitem{Schiffer1995} Schiffer, P.; Ramirez, A.P.; Bao, W.; Cheong, S.-W. Study of t$\overline{t}$ Production in p$\overline{p}$ Collisions Using Total Transverse Energy. \emph{Phys. Rev. Lett.} \textbf{1995}, \emph{75}, 3336.
\bibitem{Hwang1995} Hwang, H.Y.; Cheong, S.-W.; Radaelli, P.G.; Marezio, M.; Batlogg, B. Lattice Effects on the Magnetoresistance in Doped LaMnO$_3$. \emph{Phys. Rev. Lett.} \textbf{1995}, \emph{75}, 914.
\bibitem{Goodenough1997} Goodenough, J.B.; Zhou, J.-S. New forms of phase segregation. \emph{Nature (London)} \textbf{1997}, \emph{386}, 229.
\bibitem{Teresa1997} Teresa, J.M.D.; Ibarra, M.R.; Algarabel, P.A.; Ritter, C.; Marquina, C.; Blasco, J.G.J.; Moral, A.D.; Arnold, Z. Evidence for magnetic polarons in the magnetoresistive perovskites. \emph{Nature (London)} \textbf{1997}, \emph{386}, 256.
\bibitem{Uehara1999} Uehara, M.; Mori, S.; Chen, C.H.; Cheong, S.-W. Percolative phase separation underlies colossal magnetoresistance in
mixed-valent manganites. \emph{Nature (London)} \textbf{1999}, \emph{399}, 560.
\bibitem{Li2012} Li, H.-F.; Xiao, Y.; Schmitz, B.; Persson, J.; Schmidt, W.; Meuffels, P.; Roth, G.; Br$\ddot{\rm u}$ckel, Th. Possible magnetic-polaron-switched positive and negative magnetoresistance in the GdSi single crystals. \emph{Sci. Rep.} \textbf{2012}, \emph{2}, 750.
\bibitem{Ruddlesden1958} Ruddlesden, S.N.; Popper, P. The compound Sr$_3$Ti$_2$O$_7$ and its structure. \emph{Acta Cryst.} \textbf{1958}, \emph{11}, 54.
\bibitem{Seshadri1997} Seshadri, R.; Maigan, A.; Hervieu, M.; Nguyen, N.; Raveau, B. Complex magnetotransport in LaSr$_2$Mn$_2$O$_7$. \emph{Solid State Commun.} \textbf{1997}, \emph{101}, 453.
\bibitem{Kubota2000} Kubota, M.; Fujioka, H.; Ohoyama, K.; Moritomo, Y.; Yoshizawa, H.; Endoh, Y. Relation between crystal and magnetic structures of layered manganite La$_{2-2\emph{x}}$Sr$_{1+2\emph{x}}$Mn$_2$O$_7$ (0.30 $<$ x $<$ 0.50). \emph{J. Phys. Soc. Jpn.} \textbf{2000}, \emph{69}, 1606.
\bibitem{Li2008} Li, H.-F. \emph{Synthesis of CMR Manganites and Ordering Phenomena in Complex Transition Metal Oxides}; Forschungszentrum J$\ddot{\rm u}$lich GmbH: J$\ddot{\rm u}$lich, Germany, 2008.
\bibitem{Moritomo1996} Moritomo, Y.; Asamitsu, A.; Kuwahara, H.; Tokura, Y. Giant magnetoresistance of manganese oxides with a layered perovskite structure. \emph{Nature (London)} \textbf{1996}, \emph{380}, 141.
\bibitem{Battle1996} Battle, P.D.; Green, M.A., Laskey, N.S.; Millburn, J.E.; Radaelli, P.G.; Rosseinsky, M.J.; Sullivan, S.P.; Vente, J.F. Crystal and magnetic structures of the colossal magnetoresistance manganates. Sr$_{2-\emph{x}}$Nd$_{1+\emph{x}}$Mn$_2$O$_7$ (\emph{x}~=~0.0, 0.1). \emph{Phys. Rev. B} \textbf{1996}, \emph{54}, 15967.
\bibitem{Moritomo1998} Moritomo, Y.; Maruyama, Y.; Akimoto, T.; Nakamura, A. Layered-type antiferromagnetic state in double-layered manganites: (La$_{1-\emph{z}}$Nd$_\emph{z}$)$_{2-2\emph{x}}$Sr$_{1+2\emph{x}}$Mn$_2$O$_7$. \emph{J. Phys. Soc. Jpn.} \textbf{1998}, \emph{67}, 405.
\bibitem{Moritomo1999} Moritomo, Y.; Ohoyama, K.; Ohashi, M. Competition of interbilayer magnetic couplings in R$_{1.4}$Sr$_{1.6}$Mn$_2$O$_7$ (R~=~La$_{1-z}$Nd$_z$). \emph{Phys. Rev. B} \textbf{1999}, \emph{59}, 157.
\bibitem{Asano1997} Asano, H.; Hayakawa, J.; Matsui, M. Two-dimensional ferromagnetic ordering and magnetoresistance in the layered perovskite La$_{2-2\emph{x}}$Ca$_{1+2\emph{x}}$Mn$_2$O$_7$. \emph{Phys. Rev. B} \textbf{2000}, \emph{56}, 5396.
\bibitem{Hirota1998} Hirota, K.; Moritomo, Y.; Fujioka, H.; Kubota, M.; Yoshizawa, H.; Endoh, Y. Neutron-diffraction studies on the magnetic ordering process in the layered Mn perovskite La$_{2-2\emph{x}}$Sr$_{1+2\emph{x}}$Mn$_2$O$_7$ (\emph{x}~=~0.40, 0.45 and 0.48). \emph{J. Phys. Soc. Jpn.} \textbf{1998}, \emph{67}, 3380.
\bibitem{Perring1998} Perring, T.G.; Aeppli, G.; Kimura, T.; Tokura, Y.; Adams, M.A. Ordered stack of spin valves in a layered magnetoresistive perovskite. \emph{Phys. Rev. B} \textbf{1998}, \emph{58}, 14693.
\bibitem{Argyriou1999} Argyriou, D.N.; Mitchell, J.F.; Radaelli, P.G.; Bordallo, H.N.; Cox, D.E.; Medarde, M.; Jorgensen, J.D. Lattice effects and magnetic structure in the layered colossal magnetoresistance manganite
La$_{2-2\emph{x}}$Sr$_{1+2\emph{x}}$Mn$_2$O$_7$, \emph{x}~=~0.3. \emph{Phys. Rev. B} \textbf{1999}, \emph{59}, 8695.
\bibitem{Kimura2000} Kimura, T.; Tokura, Y. Layered magnetic manganites. \emph{Annu. Rev. Mater. Sci.} \textbf{2000}, \emph{30}, 451.
\bibitem{Qiu2004} Qiu, X.; Billinge, S.J.L.; Kmety, C.R.; Mitchell, J.F. Evidence for nano-scale inhomogeneities in bilayer manganites in the Mn$^{4+}$ rich region: $0.54 \leq x \leq 0.80$. \emph{J. Phys. Chem. Solids} \textbf{2004}, \emph{65}, 1423.
\bibitem{Zheng2008} Zheng, H.; Li, Q.A.; Gray, K.E.; Mitchell, J.F. Charge and orbital ordered phases of La$_{2-2x}$Sr$_{1+2x}$Mn$_2$O$_{7-\delta}$. \emph{Phys. Rev. B} \textbf{2008}, \emph{78}, 155103.
\bibitem{Sonomura2013} Sonomura, H.; Terai, T.; Kakeshita, T.; Osakabe, T.; Kakurai, K. Neutron diffraction study on magnetic structures in a La$_{1.37}$Sr$_{1.63}$Mn$_2$O$_7$ single crystal under hydrostatic pressures of up to 0.8 GPa. \emph{Phys. Rev. B} \textbf{2013}, \emph{87}, 184419.
\bibitem{Guptasarma2005} Guptasarma, P.; Williamsen, M.S.; Ray, S.K. Floating zone growth of bulk single crystals of complex oxides. \emph{Mater. Res. Soc. Symp. Proc.} \textbf{2005}, \emph{848}.
\bibitem{Ito2013} Ito, T.; Ushiyama, T.; Yanagisawa, Y.; Tomioka, Y.; Shindo, I.; Yanase, A. Laser-diode-heated floating zone (LDFZ) method appropriate to crystal growth of incongruently melting materials. \emph{J. Cryst. Growth} \textbf{2013}, \emph{363}, 264.
\bibitem{Wu2020} Wu, S.; Zhu, Y.H.; Gao, H.S.; Xiao, Y.G.; Xia, J.C.; Zhou, P.F.; Ouyang, D.F.; Li, Z.; Chen, Z.Q.; Tang, Z.K.; et al. Super-necking crystal growth and structural and magnetic properties of SrTb$_2$O$_4$ single crystals. \emph{ACS Omega} \textbf{2020}, \emph{5}, 16584--16594.
\bibitem{Ouladdiaf2006} Ouladdiaf, B.; Archer, J.; McIntyre, G.J.; Hewat, A.W.; Brau, D.; York, S. OrientExpress: A new system for Laue neutron diffraction. \emph{Phys. B. Condens. Matter} \textbf{2006}, \emph{385--386}, 1052.
\bibitem{Fullprof} Rodr{\'{i}}guez-Carvajal, J. Recent advances in magnetic-structure determination by neutron powder diffraction. \emph{Phys. B.  Condens. Matter} \textbf{1993}, \emph{192}, 55.
\bibitem{Li2007-2} Li, H.-F.; Su, Y.; Persson, J.; Meuffels, P.; Walter, J.M.; Skowronek, R.; Br$\textrm{\"{u}}$ckel, T. Neutron-diffraction study of structural transition and magnetic order in orthorhombic and rhombohedral La$_{7/8}$Sr$_{1/8}$Mn$_{1-\gamma}$O$_{3+\delta}$. \emph{ J. Phys. Condens. Matter} \textbf{2007}, \emph{19}, 176226.
\bibitem{Zhu2020} Zhu, Y.H.; Wu, S.; Jin, S.J.; Huq, A.; Persson, J.; Gao, H.S.; Ouyang, D.F.; He, Z.B.; Yao, D.-X.; Tang, Z.K.; et al. High-temperature magnetism and crystallography of a YCrO$_3$ single crystal. \emph{Phys. Rev. B} \textbf{2020}, \emph{101}, 014114.

\end{thebibliography}
\end{document}